\documentclass[showpacs,fleqn,nobibnotes]{revtex4}

\usepackage{amsmath}
\usepackage{graphicx}

\def\lsim{\raise0.3ex\hbox{$<$\kern-0.75em\raise-1.1ex\hbox{$\sim$}}}
\def\gsim{\raise0.3ex\hbox{$>$\kern-0.75em\raise-1.1ex\hbox{$\sim$}}}

\newcommand{\rk}{\mbox{\boldmath $k$}}

\newcommand{\rr}{\mbox{\boldmath $r$}}

\newcommand{\rb}{\mbox{\boldmath $b$}}

\begin{document}

\title{Exclusive processes in electron - ion collisions}
\pacs{12.38.-t, 24.85.+p, 25.30.-c}
\author{E. R. Cazaroto$^a$, F. Carvalho$^b$, V.P. Gon\c{c}alves$^c$, 
M.S. Kugeratski$^d$ and F. S. Navarra$^a$}

\affiliation{
$^a$ Instituto de F\'{\i}sica, Universidade de S\~{a}o Paulo,
C.P. 66318,  05315-970 S\~{a}o Paulo, SP, Brazil\\
$^b$ Departamento de Ci\^encias Exatas e da Terra, Universidade Federal de S\~ao Paulo,\\  
Campus Diadema, Rua Prof. Artur Riedel, 275, Jd. Eldorado, 09972-270, Diadema, SP, Brazil\\
$^c$ Instituto de F\'{\i}sica e Matem\'atica, Universidade Federal de Pelotas\\
Caixa Postal 354,  96010-900, Pelotas, RS, Brazil.\\
$^d$ Centro de Engenharia da Mobilidade, Universidade Federal de Santa Catarina,\\
Campus Universit\'ario, Bairro Bom Retiro  89219-905, Joinville, SC, Brazil.\\
}

\begin{abstract}
The exclusive processes in electron-ion ($eA$) interactions  are an important tool to 
investigate the QCD dynamics at high energies as they are in general driven by the gluon 
content of the target which is strongly subject to parton saturation effects.  In this 
paper we compute the cross sections for the exclusive vector meson production as well as 
the deeply virtual Compton scattering (DVCS) relying on the color dipole approach and 
considering the numerical solution of the  Balitsky-Kovchegov  equation including running 
coupling corrections (rcBK). The production cross sections obtained with the rcBK solution  and bCGC parametrization are very similar, the former being slightly larger. 

\end{abstract}

\maketitle

Exclusive processes in deep inelastic scattering (DIS) have appeared as key reactions to 
trigger the generic mechanism of diffractive scattering (For a recent review see, e.g. 
\cite{scho}). In particular, diffractive vector meson production and deeply virtual Compton 
scattering (DVCS) have been extensively studied at HERA and provide a valuable  probe of the  
QCD dynamics at high energies. These processes are driven by the gluon 
content of the target (proton or nucleus) which is strongly subject to parton saturation 
effects as well as to  nuclear shadowing corrections when one considers scattering 
on nuclei (See e.g. \cite{kope_dvcs}). In particular, the cross sections for exclusive processes in DIS are proportional 
to the square of the scattering amplitude, which makes them  strongly sensitive to the 
underlying QCD dynamics. In a recent paper \cite{vmprc} we have estimated the coherent and 
incoherent cross sections for  exclusive $\rho$ and $J/\Psi$ production considering the 
color dipole approach and phenomenological saturation models which describe the scarce 
$F_2^A$ data as well as the HERA data. Our results demonstrated that the coherent production 
of vector mesons is dominant, with a small contribution coming from incoherent processes. 
Moreover, our results indicate that the experimental study of these processes is feasible 
in future electron - ion collider, as e.g. the eRHIC \cite{raju_ea2} or LHeC \cite{dainton}.
In this paper we complement our previous analysis including $\phi$ production and extending 
our study to the nuclear DVCS (See also \cite{mag}). Moreover, we review our results for 
$\rho$ and $J/\Psi$ production making use of the numerical solution of the  
Balitsky-Kovchegov equation \cite{BAL,KOVCHEGOV} including running coupling corrections 
\cite{kovwei1,javier_kov,balnlo} in order to estimate the contribution of the saturation 
physics to exclusive processes. 
Our main motivation is associated to the fact that the improved BK equation has been shown 
to be really successful when applied to the description of the $ep$ HERA data on inclusive 
and diffractive proton structure function \cite{bkrunning,weigert,vic_joao}, as well as on 
exclusive processes \cite{vicmagane} and on the  forward hadron spectra in $pp$ and $dA$ 
collisions \cite{vic_joao,alba_marquet}.

Let us start presenting a brief review of exclusive processes in electron - ion collisions 
(For details see \cite{vmprc,kop1}). In the color dipole approach the exclusive  production $\gamma^* A \rightarrow EY$  ($E = \rho, \phi, J/\Psi$ or $\gamma$) in electron-nucleus interactions at high energies can be 
factorized in terms of the fluctuation of the virtual photon into a $q \bar{q}$ color 
dipole, the dipole-nucleus scattering by a color singlet exchange  and the recombination into 
the exclusive final state $E$. This process is characterized by a rapidity gap in the final 
state. If the nucleus scatters elastically, $Y = A$, the process is called coherent 
production and the corresponding integrated cross 
section  is given in the high energy regime (large coherence length: $l_c \gg R_A$) by 
\cite{vmprc,kop1}
\begin{eqnarray}
\sigma^{coh}\, (\gamma^* A \rightarrow EA)  =  \int d^2\rb \left\langle 
\mathcal{N}^A(x,\rr,\rb) \right\rangle^2
\label{totalcscoe}
\end{eqnarray}
where
\begin{eqnarray}
\left\langle \mathcal{N} \right\rangle = \int d^2\rr
 \int dz  \Psi_E^*(\rr,z) \, \mathcal{N}^A(x,\rr,\rb)\, \Psi_{\gamma^*}(\rr,z,Q^2)
 \label{totalcscoe1}
\end{eqnarray}
and $ {\cal N} (x, \rr, \rb)$ is the forward dipole-target scattering amplitude for a 
dipole with 
size $\rr$ and impact parameter $\rb$ which encodes all the
information about the hadronic scattering, and thus about the
non-linear and quantum effects in the hadron wave function. 
On the other hand, if the nucleus scatters inelastically, i.e. breaks up ($Y = X$),   
the process is denoted incoherent production. 
 In this case 
one sums over all final states of the target nucleus, 
except those that contain particle production. The $t$ slope is the same as in the case of a 
nucleon target. Therefore we have: 
\begin{eqnarray}
\sigma^{inc}\, (\gamma^* A \rightarrow EX) = \frac{|{\cal I}m \, 
{\cal A}(s,\,t=0)|^2}{16\pi\,B_E} \;
\label{totalcsinc}
\end{eqnarray}
where at high energies ($l_c \gg R_A$) \cite{kop1}:
\begin{eqnarray}
|{\cal I}m \, {\cal A}|^2  =  \int d^2\rb \, T_A(\rb) \left\langle  \sigma_{dp} \, 
\exp[- \frac{1}{2} \, \sigma_{dp} \, T_A(\rb)]  \right\rangle^2 
\label{totalcsinc1}
\end{eqnarray} 
and $ \sigma_{dp}$ is the dipole-proton cross section, which in the eikonal approximation it 
is  given by:
\begin{equation} 
\sigma_{dp} (x, \rr) = 2 \int d^2 \rb \,  {\cal N}^p (x, \rr, \rb)\,\,.
\label{sdip}
\end{equation}
In the incoherent case, the $q\bar{q}$ pair attenuates with a constant absorption cross 
section, as in the Glauber model, except that the whole exponential is averaged 
rather than just the cross section in the exponent.  
As discussed in \cite{vmprc}, the coherent and incoherent cross sections depend  
differently on  $t$.  
At small-$t$ ($-t\,R_A^2/3 \ll 1$) coherent 
production dominates, with the signature being a sharp 
forward diffraction peak. On the other hand, incoherent production will dominate  
at large-$t$ ($-t\,R_A^2/3 \gg 1$), with the $t$-dependence being to a good 
accuracy the same as in the production off  free nucleons.

In the Eqs. (\ref{totalcscoe1}) and (\ref{totalcsinc1}) the functions 
$\Psi^{\gamma}(z,\,\rr)$ and $\Psi^{E}(z,\,\rr)$  
are the light-cone wavefunctions  of the photon and the exclusive final state, respectively.  The 
variable $\rr$ defines the relative transverse
separation of the pair (dipole) and $z$ $(1-z)$ is the
longitudinal momentum fraction of the quark (antiquark). 
In the dipole formalism, the light-cone
 wavefunctions $\Psi(z,\,\rr)$ in the mixed
 representation $(r,z)$ are obtained through a two dimensional Fourier
 transform of the momentum space light-cone wavefunctions
 $\Psi(z,\,\rk)$.
The photon wavefunctions  are well known in literature \cite{KMW}. For the meson 
wavefunction, we have considered the Gauss-LC  model of Ref. \cite{KMW}.  
The motivation for this choice is its simplicity and the fact 
that the results are not very sensitive to differences between the models analyzed in \cite{KMW}.  We choose 
the quark masses to be $m_{u,d,s} = 0.14$ GeV and $m_c = 1.4$ GeV. The parameters 
for the meson wavefunction can be found in Ref. \cite{KMW}. In the DVCS case, as one has a 
real photon at the final state, only the transversely polarized overlap function contributes 
to the cross section.  Summed over the quark helicities, for a given quark flavour $f$ it is 
given by \cite{MW},
\begin{eqnarray}
  (\Psi_{\gamma}^*\Psi)_{T}^f & = & \frac{N_c\,\alpha_{\mathrm{em}}
e_f^2}{2\pi^2}\left\{\left[z^2+\bar{z}^2\right]\varepsilon_1 K_1(\varepsilon_1 r) 
\varepsilon_2 K_1(\varepsilon_2 r) 
 +     m_f^2 K_0(\varepsilon_1 r) K_0(\varepsilon_2 r)\right\},
  \label{eq:overlap_dvcs}
\end{eqnarray}
where we have defined the quantities $\varepsilon_{1,2}^2 = z\bar{z}\,Q_{1,2}^2+m_f^2$ and 
$\bar{z}=(1-z)$. Accordingly, the photon virtualities are $Q_1^2=Q^2$ (incoming virtual 
photon) and $Q_2^2=0$ (outgoing real photon).

\begin{figure}[t]
\includegraphics[scale=0.50]{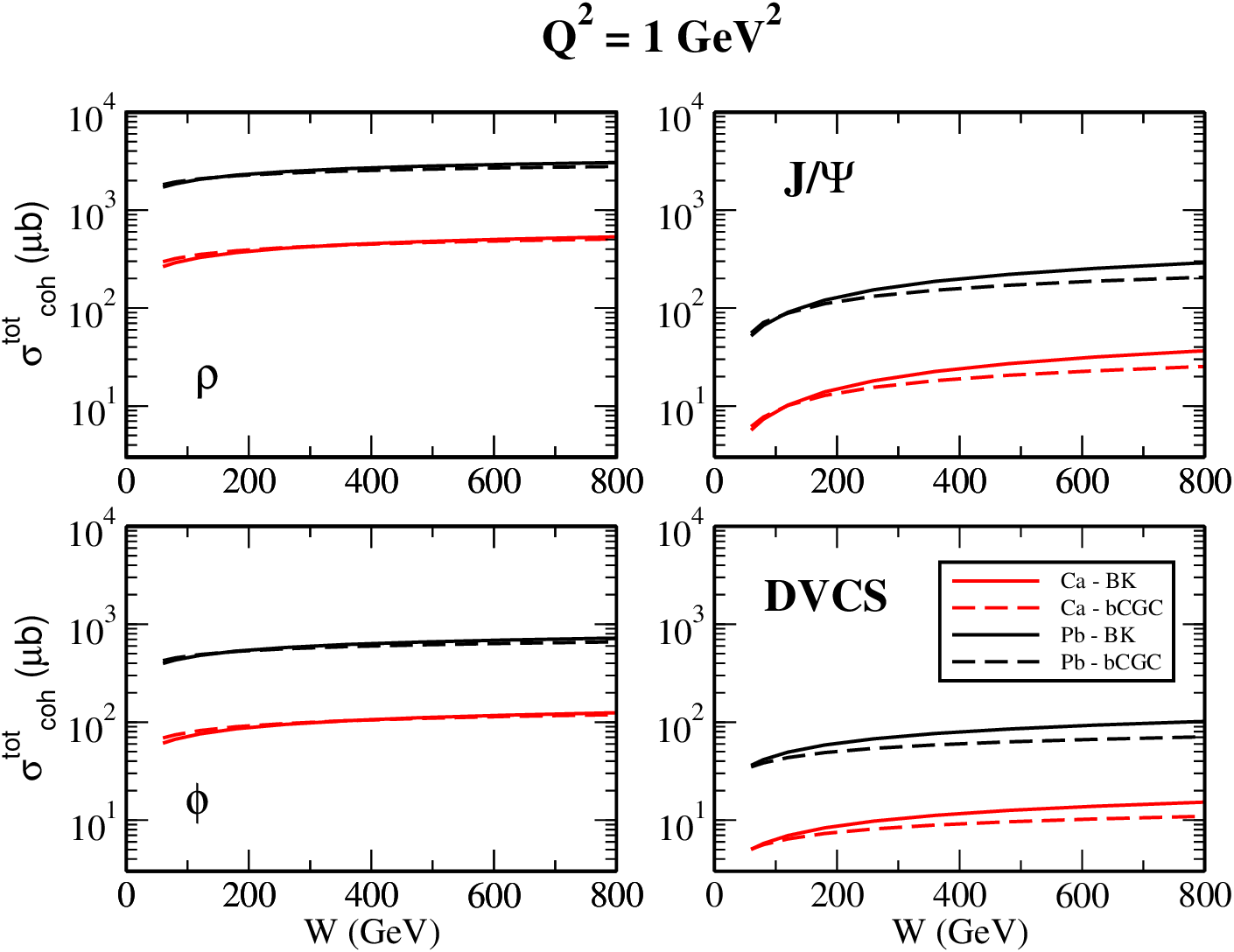}
\vspace{0.4cm}
\caption{(Color online)  Energy dependence of the coherent cross section at different 
final states and $Q^2 = 1$ GeV$^2$.}
\label{fig:1}
\end{figure}

In order to estimate the coherent production in $eA$ collisions we need to specify the 
forward dipole - nucleus scattering amplitude, $\mathcal{N}^A(x,\rr,\rb)$.  
Following \cite{vmprc} we will use in our calculations  the model proposed in Ref. 
\cite{armesto}, which describes  the current  experimental data on the nuclear 
structure function as well as includes the  impact parameter dependence in the dipole 
nucleus cross section. In this model the forward dipole-nucleus amplitude is given by
\begin{eqnarray}
{\cal{N}}^A(x,\rr,\rb) = 1 - \exp \left[-\frac{1}{2}  \, \sigma_{dp}(x,\rr^2) 
\,T_A(\rb)\right] \,\,,
\label{enenuc}
\end{eqnarray}
where $\sigma_{dp}$ is the dipole-proton cross section and $T_A(\rb)$ is the nuclear profile 
function, which is obtained from a 3-parameter Fermi distribution for the nuclear
density normalized to $A$.
The above equation
sums up all the 
multiple elastic rescattering diagrams of the $q \overline{q}$ pair
and is justified for large coherence length, where the transverse separation $\rr$ of partons 
in the multiparton Fock state of the photon becomes a conserved quantity, {\it i.e.} 
the size 
of the pair $\rr$ becomes eigenvalue
of the scattering matrix.

In our approach  the coherent [Eq. (\ref{totalcscoe})] and 
incoherent [Eq. (\ref{totalcsinc})] cross sections can be calculated in terms of the 
dipole-proton cross section or the forward dipole-proton scattering amplitude 
[See Eq. (\ref{sdip})], which is a solution of the BK equation. As the leading order 
solution of the BK equation was not able to describe the HERA data, in Ref. \cite{vmprc} 
we have used  the GBW \cite{GBW} and bCGC \cite{KMW}
parametrizations for ${\cal N}^p $ as input in our calculations. However, in the last years 
the next-to-leading order corrections to the  BK equation were calculated  
\cite{kovwei1,javier_kov,balnlo}. Such calculation allows one to estimate 
the soft gluon emission and running coupling corrections to the evolution kernel.
The authors have verified that  the dominant contributions come from the running 
coupling corrections, which allow us to  determine the scale of the running coupling in the 
kernel. The solution of the improved BK equation was studied in detail in Refs. 
\cite{javier_kov,javier_prl}. Basically, one finds that the running of the coupling reduces 
the speed of the evolution to values compatible with experimental data. In \cite{bkrunning} 
a global analysis of the small $x$ data for the proton structure function using the improved BK 
equation was performed  (See also Ref. \cite{weigert}). In contrast to the  BK  equation 
at leading logarithmic $\alpha_s \ln (1/x)$ approximation, which  fails to describe the HERA 
data, the inclusion of running coupling effects in the evolution renders the BK equation 
compatible with them (See also \cite{vic_joao,alba_marquet,vicmagane}).
It is important to emphasize that the impact parameter dependence was not taken into account 
in Ref. \cite{bkrunning}, the normalization of the dipole cross section was fitted to data 
and two distinct initial conditions, inspired by  the Golec Biernat-Wusthoff (GBW) \cite{GBW} 
and McLerran-Venugopalan (MV) \cite{MV} models, were considered. The predictions resulted 
to be almost independent of the initial conditions and, besides, it was observed that it 
is impossible to describe the experimental data using only the linear limit of the BK 
equation. The parametrizations obtained in \cite{bkrunning} were very successful in 
reproducing DIS data but it remains to seen whether they  can also be used to describe data 
from RHIC. Other parametrizations of dipole cross sections had to be slightly modified in 
order to account for RHIC data \cite{nos2006}.

In what follows we calculate the exclusive observables using as input in our calculations 
the solution of the running coupling Balitsky-Kovchegov (rcBK) evolution equation. We make
use of the public-use code available in \cite{code}. In numerical calculations we have 
considered the GBW initial condition for the evolution. Furthermore, we compare the rcBK 
predictions with those from the bCGC model used in our previous calculation \cite{vmprc}.
Moreover, in order to calculate the incoherent cross section for vector meson production we 
will use the following  parametrization of the diffractive slope
\begin{eqnarray}
B_V\,(Q^2) = 0.60\,\left[ \frac{14}{(Q^2+M_V^2)^{0.26}}+1  \right]                
\end{eqnarray}                                         
obtained from a fit to experimental data referred in Ref. \cite{Mara}.
In the DVCS case we take the experimental parametrization \cite{H1_dvcs}, 
$B\,(Q^2)=a[1-b\log(Q^2/Q_0^2)]$, with $a=6.98 \pm 0.54 $ GeV$^2$, $b=0.12 \pm 0.03$ and 
$Q_0^2 = 2$ GeV$^2$.

\begin{figure}[t]
\includegraphics[scale=0.50]{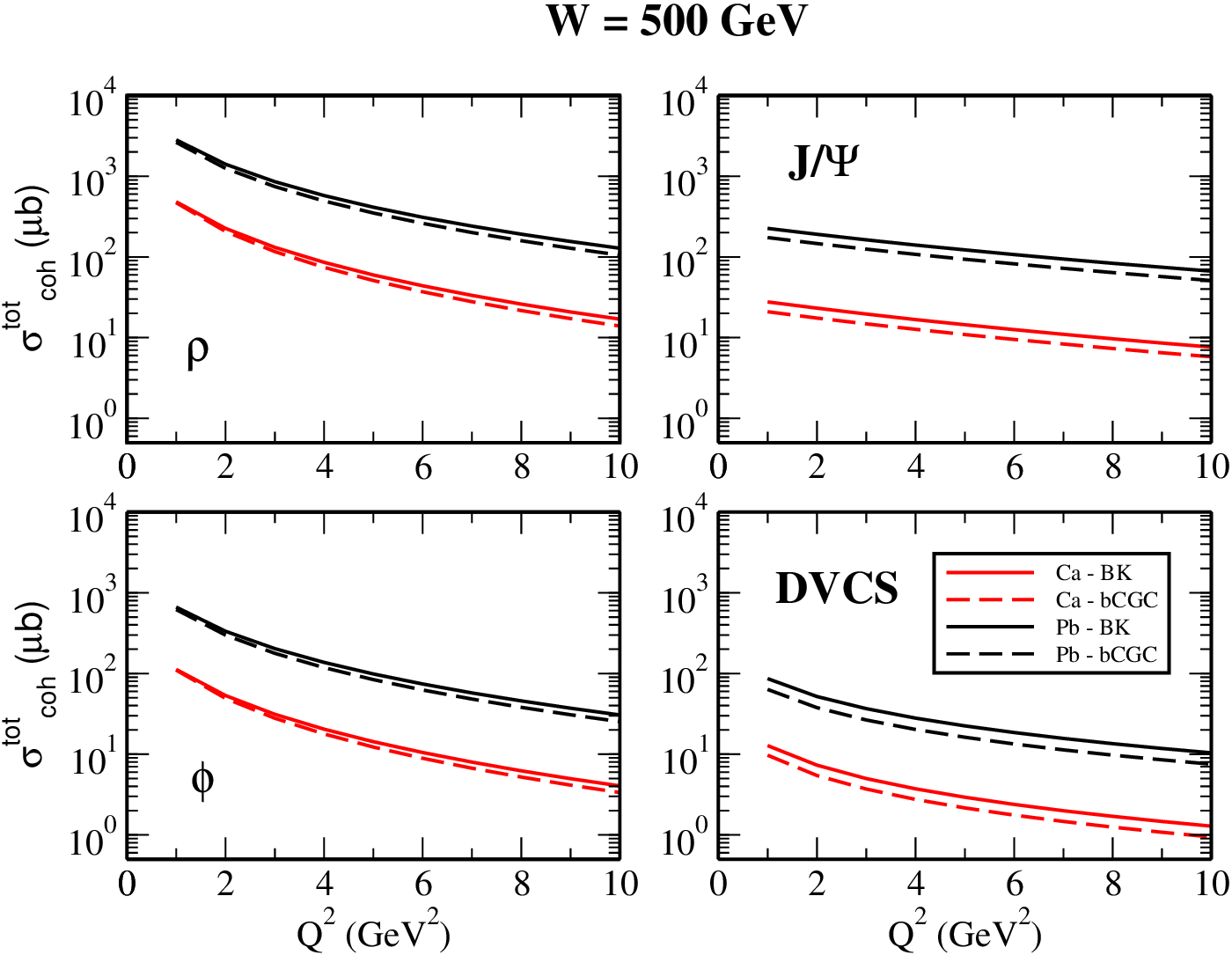}
\vspace{0.4cm}
\caption{(Color online) Dependence on the photon virtuality of the coherent cross section 
for different final states and $W = 500$ GeV.}
\label{fig:2}
\end{figure}

\begin{figure}[t]
\includegraphics[scale=0.50]{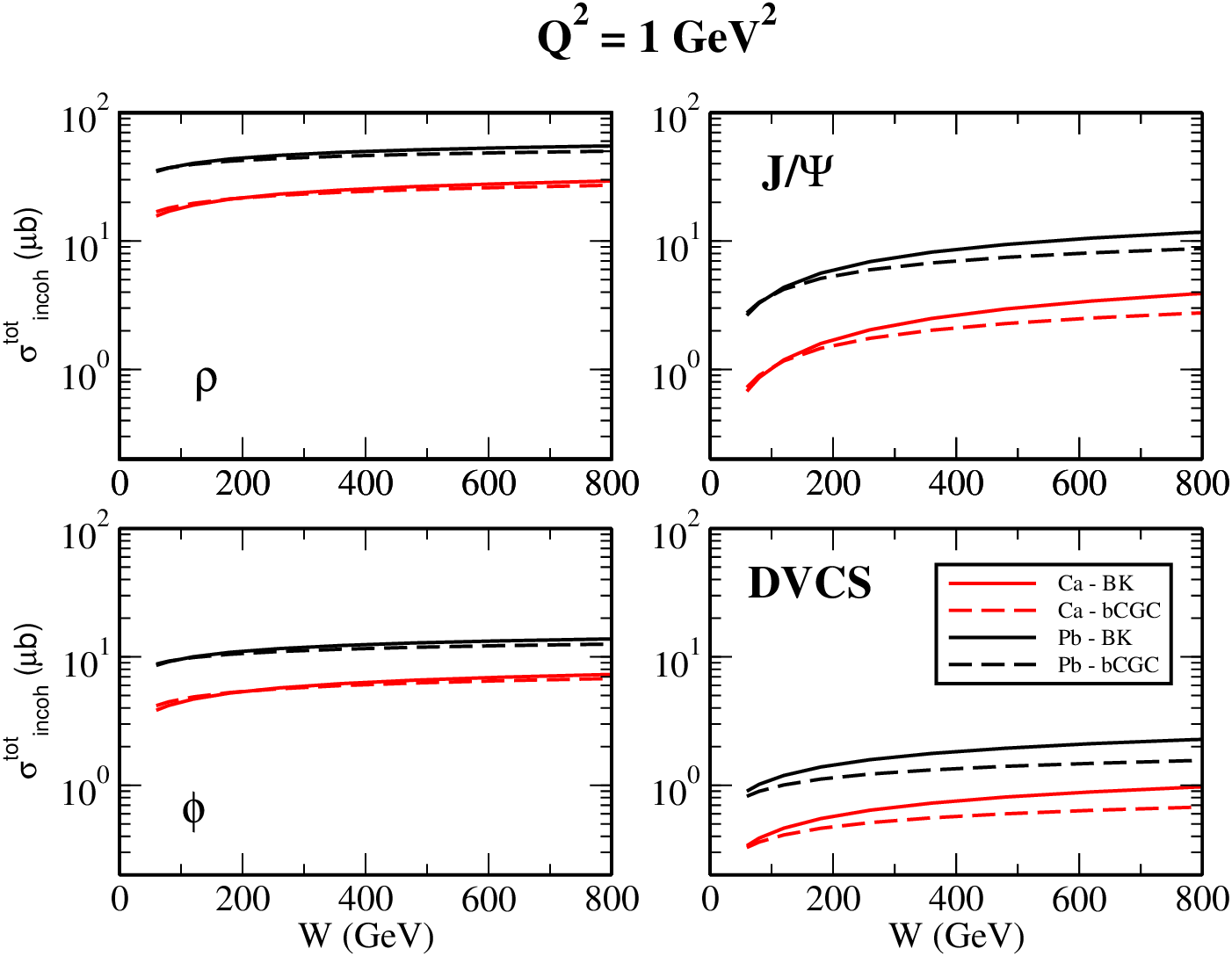}
\vspace{0.4cm}
\caption{(Color online)  Energy dependence of the incoherent cross section for different 
final states and $Q^2 = 1$ GeV$^2$.}
\label{fig:3}
\end{figure}

\begin{figure}[t]
\includegraphics[scale=0.50]{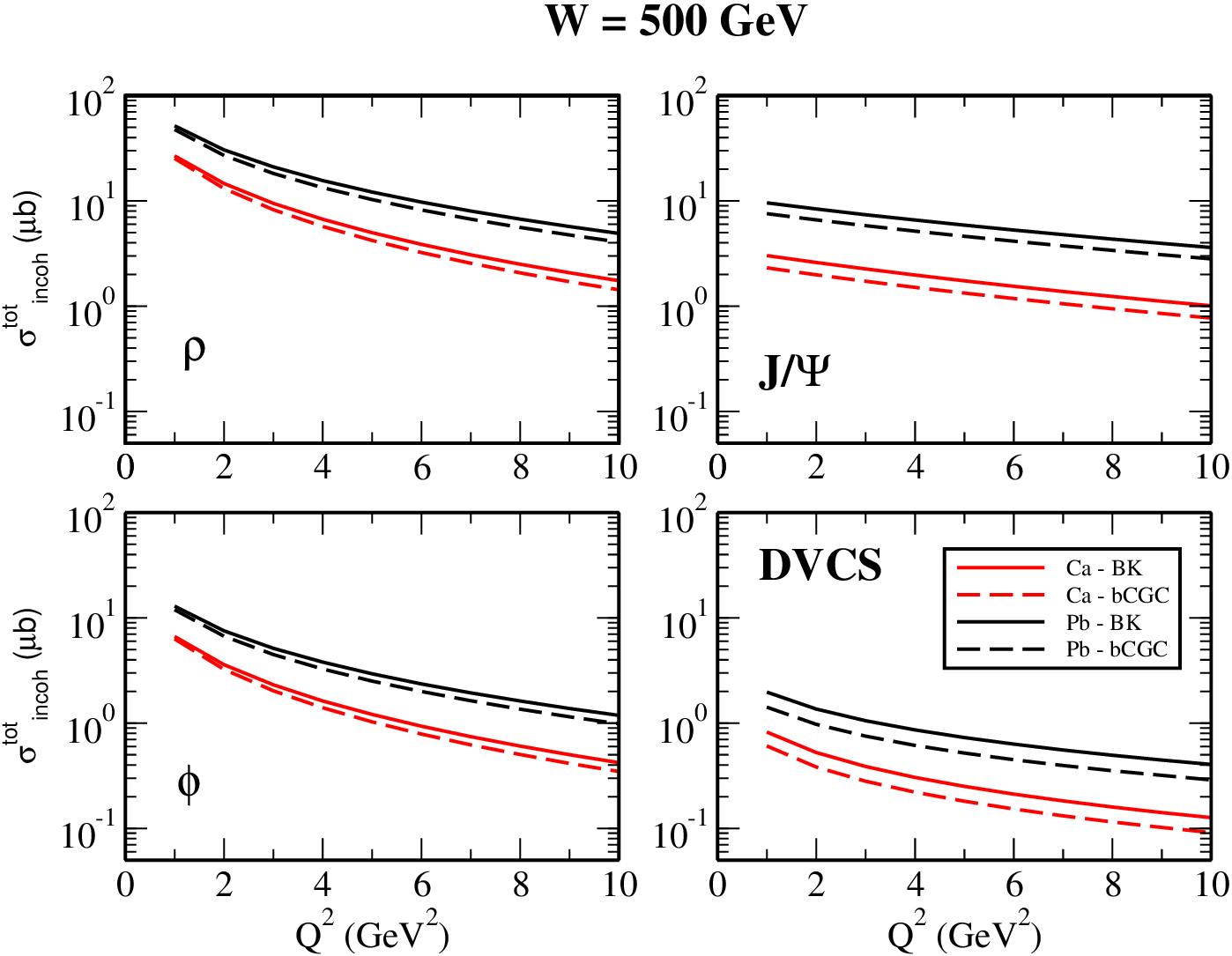}
\vspace{0.4cm}
\caption{(Color online) Dependence on the photon virtuality of the incoherent cross 
section for different final states and $W = 500$ GeV.}
\label{fig:4}
\end{figure}

In Fig. \ref{fig:1} we show the coherent production cross section as a function of the photon-target 
c.m.s energy, $W$, for a fixed photon virtuality $Q^2=1$ GeV$^2$. Each one of the panels 
shows  the results obtained for one specific  final state. In each 
single figure  the two upper (lower) curves show the results for a Pb (Ca) target.  
In  all figures the 
dashed (solid) lines are obtained with the bCGC (rcBK) dipole cross section. Fig. \ref{fig:2} shows
the same cross sections, this time as a function of $Q^2$ for a fixed energy, $W=500$ GeV. 
Figs. \ref{fig:3} and \ref{fig:4} are the exact analogues (of Figs. \ref{fig:1} and \ref{fig:2}) for the corresponding incoherent 
cross sections.

The curves in the figures have the merit of being the first concrete predictions made for 
these processes with the help of the recently obtained rcBK dipole cross section.  They 
present some features which are expected  and some other features which could not have 
been anticipated without a quantitative calculation. In first place we observe, as it 
should be, that all cross sections grow with $W$ and fall with $Q^2$. The first feature 
is related solely to the  nature of the dipole cross section, which grows with the 
energy, whereas the second feature comes from the dipole wave functions. We can also see 
from the figures that, at least for the two cases considered (bCGC and rcBK), the production 
cross sections are not very strongly  dependent on  the choice of the dipole cross section.

At low $Q^2$ and low  $W$ the  bCGC and rcBK  production cross sections are indistinguishable 
one from the other because the  dipole cross sections tend to coincide. These latter have been 
tuned to fit DIS data,  which are taken in this kinematical region. Another expected feature 
is the observed decrease of the cross sections with increasing  
vector meson masses, which comes from the wave functions.

Differences are expected to appear at higher energies, where we enter  
the  lower $x$ (extrapolation)  region. In all cases we see that  the  results obtained 
with the rcBK cross section are larger  than those obtained with the bCGC one. This is 
related to the fact that the numerical solutions of the BK equation tend to reach later 
the unitarity limit \cite{marcosandre}. In the first estimates, with a fixed coupling, 
the solutions of the BK equation would saturate too fast. In subsequent studies it was found 
that running coupling corrections to the BK kernel could bring the evolution speed down to 
values compatible with those extracted from data, but still larger than those found in other 
parametrizations, such as the bCGC one. Due to this fact, the
results obtained with the rcBK dipole cross section grow faster with energy than those 
obtained with the bCGC one. 

A curious feature in the figures  is that {\it the differences between bCGC and rcBK are 
larger for heavier vector mesons}. This can be 
understood looking carefully at the integrand of   (\ref{totalcscoe}), which is the 
product of 
the  wave functions, containing information about the masses, and the dipole cross section.
As a function of the dipole size $r$ the difference between bCGC and rcBK is mostly in the 
low to intermediate $r$ region, where the bCGC is always below the rcBK dipole cross 
section. At large $r$ the two cross sections are close to each other. The overlap function, 
i.e., $ (\Psi_{\gamma}^*\Psi)^f$ given by  (\ref{eq:overlap_dvcs}) 
(with the inclusion of the longitudinally polarized overlap function), has peaks at 
different locations.  The $\rho$ is a larger state  and its overlap function peaks at 
much larger values than the $J/\psi$ overlap function. In this way it gives a stronger  
weight to larger $r$ where the differences between bCGC and rcBK are smaller. The same 
thing  happens to the $\phi$. On the other hand, the $J/\psi$ overlap function peaks at 
smaller $r$ where the dipole cross sections are more different from each other. A similar 
behavior is verified in the DVCS case.  

As shown in \cite{marcosandre}, at increasing values of the energy $W$ 
(and thus of smaller of $x$) the difference between rcBK and bCGC moves to smaller values 
of $r$, a region which is suppressed by the overlap functions of the $\rho$ and $\phi$. 
This explains why the $\rho$ (and also the $\phi$) production cross sections are the 
almost the same for rcBK and bCGC dipole cross sections for all the energies considered, 
as it can be seen on the right side of Figs. 1 and 3. 

In Figs. 2 and 4 we would expect to see a convergence of the curves for higher values 
of $Q^2$. In this region the dipoles are small and all cross sections should approach the
color transparency regime. In fact, a difference between them persists even at large 
$Q^2$ because in the expressions used here there is no DGLAP evolution, which would bring
the dipole cross sections together.

As a summary, we presented a systematic analysis of exclusive production in small-$x$ 
deep inelastic electron - ion scattering in terms of the non-linear QCD dynamics. This 
was the first calculation (of these observables)  using the  solution of the BK equation  
improved with running coupling corrections. In this work we obtain  predictions for the 
exclusive production of vector mesons and DVCS.  Our analysis confirms the dominance of 
the coherent production  with a small contribution coming from incoherent processes, a 
result previously found in  \cite{vmprc}. Our main result is that the BK evolution 
equation implies larger cross sections for exclusive processes than  the phenomenological 
model proposed in \cite{KMW}, the so called  bCGC model. Our predictions for both vector 
meson and DVCS production are relevant for the physics programs of the future experiments 
eRHIC and LHeC.

\begin{acknowledgments}
This work was  partially financed by the Brazilian funding
agencies CNPq and FAPESP.
\end{acknowledgments}

\begin{thebibliography}{99}

\bibitem{scho}
  L.~Schoeffel,
 Prog.\ Part.\ Nucl.\ Phys.\  {\bf 65}, 9 (2010).

\bibitem{kope_dvcs}
B.~Z.~Kopeliovich, I.~Schmidt and M.~Siddikov,
  Phys.\ Rev.\  D {\bf 81}, 094013 (2010); Phys.\ Rev.\  D {\bf 82}, 014017 (2010).



\bibitem{vmprc}
  V.~P.~Goncalves, M.~S.~Kugeratski, M.~V.~T.~Machado and F.~S.~Navarra,
  Phys.\ Rev.\  C {\bf 80}, 025202 (2009).

\bibitem{raju_ea2}
A.~Deshpande, R.~Milner, R.~Venugopalan and W.~Vogelsang,
  Ann.\ Rev.\ Nucl.\ Part.\ Sci.\  {\bf 55},  165 (2005).


\bibitem{dainton}
  J.~B.~Dainton, M.~Klein, P.~Newman, E.~Perez and F.~Willeke,
  JINST {\bf 1}, P10001 (2006).

\bibitem{mag}
  M.~V.~T.~Machado,
  Eur.\ Phys.\ J.\  C {\bf 59}, 769 (2009).



\bibitem{BAL}  I. I. Balitsky,  Phys. Rev. Lett. {\bf 81}, 2024 (1998); 
               Phys. Lett. B  {\bf 518}, 235 (2001);   
               I.I. Balitsky and  A.V. Belitsky, Nucl. Phys. B {\bf 629}, 290  (2002). 

\bibitem{KOVCHEGOV} Y.V. Kovchegov,  Phys. Rev. D {\bf 60},  034008 (1999);  
                    Phys. Rev. D {\bf 61} 074018 (2000). 


\bibitem{kovwei1}  Y.~V.~Kovchegov and H.~Weigert, 
                   Nucl.\ Phys.\  A {\bf 784}, 188 (2007); 
                  Nucl.\ Phys.\  A {\bf 789}, 260 (2007);   
                  Y.~V.~Kovchegov, J.~Kuokkanen, K.~Rummukainen and H.~Weigert, 
                  Nucl.\ Phys.\  A {\bf 823}, 47 (2009).




\bibitem{javier_kov}  J.~L.~Albacete and Y.~V.~Kovchegov,  
                      Phys.\ Rev.\  D {\bf 75}, 125021 (2007).

\bibitem{balnlo}   I.~Balitsky,  Phys.\ Rev.\  D {\bf 75}, 014001 (2007); 
                   I.~Balitsky and G.~A.~Chirilli, 
                   Phys.\ Rev.\  D {\bf 77}, 014019 (2008).



\bibitem{bkrunning}  J.~L.~Albacete, N.~Armesto, J.~G.~Milhano and C.~A.~Salgado,
  Phys. Rev. D {\bf 80}, 034031 (2009). 


\bibitem{weigert}   H.~Weigert, J.~Kuokkanen and K.~Rummukainen,
  AIP Conf.\ Proc.\  {\bf 1105}, 394 (2009).

\bibitem{vic_joao}
  M.~A.~Betemps, V.~P.~Goncalves and J.~T.~de Santana Amaral,
 Eur.\ Phys.\ J.\  C {\bf 66}, 137 (2010).
  
\bibitem{vicmagane}
  V.~P.~Goncalves, M.~V.~T.~Machado and A.~R.~Meneses,
  Eur.\ Phys.\ J.\  C {\bf 68}, 133 (2010).  


\bibitem{alba_marquet}   J.~L.~Albacete and C.~Marquet, 
                           Phys.\ Lett.\  B {\bf 687}, 174 (2010).
  


\bibitem{kop1}   B.~Z.~Kopeliovich, J.~Nemchik, A.~Schafer and A.~V.~Tarasov, 
                 Phys.\ Rev.\  C {\bf 65}, 035201  (2002).



\bibitem{KMW}  H.~Kowalski, L.~Motyka and G.~Watt, 
               Phys.\ Rev.\  D {\bf 74}, 074016 (2006).

\bibitem{MW}  M.~Wusthoff and A.~D.~Martin, 
              J.\ Phys.\ G {\bf 25}, R309 (1999). 


\bibitem{armesto}  N.~Armesto,  Eur.\ Phys.\ J.\  C {\bf 26}, 35 (2002).




\bibitem{GBW} K. Golec-Biernat and  M. W\"usthoff,  Phys. Rev. D {\bf 59}, 014017 (1999),
{\it ibid.} D  {\bf  60},  114023 (1999).



\bibitem{javier_prl}  J.~L.~Albacete,
  Phys.\ Rev.\ Lett.\  {\bf 99}, 262301 (2007). 



\bibitem{MV} L. McLerran and R. Venugopalan,  Phys.\ Rev.\ D {\bf 49},  2233  (1994).


\bibitem{nos2006} V.~P.~Goncalves, M.~S.~Kugeratski, M.~V.~T.~Machado and F.~S.~Navarra, 
                  Phys.\ Lett.\  B {\bf 643}, 273 (2006).

\bibitem{code} http://www-fp.usc.es/phenom/rcbk


\bibitem{Mara} A.~C.~Caldwell and M.~S.~Soares,
              Nucl.\ Phys.\ A {\bf 696}, 125 (2001).


\bibitem{H1_dvcs}   F.~D.~Aaron {\it et al.}  [H1 Collaboration],
  Phys.\ Lett.\  B {\bf 659}, 796 (2008).

\bibitem{marcosandre} A systematic comparison of the energy behavior of  
                      different dipole amplitudes (including the rcBK one) 
                      can be found in: 
                      M.~A.~Betemps, V.~P.~Goncalves and J.~T.~de Santana Amaral,
  Eur.\ Phys.\ J.\  C {\bf 66}, 137 (2010).
    



\end{thebibliography}
\end{document}